\documentclass[superscriptaddress,twocolumn,showpacs,prb,floatfix]{revtex4}
\usepackage{epsfig}

\begin{document}

\title{Dynamic critical behaviour in Ising spin glasses}

\author{Michel Pleimling}
\affiliation{Institut f\"ur Theoretische Physik I, Universit\"at Erlangen-N\"urnberg, D-91058 Erlangen, Germany}

\author{I.~A.~Campbell}
\affiliation{Laboratoire des Collo\"{i}des, Verres et Nanomat\'eriaux, Universit\'e Montpellier II,
34095 Montpellier, France}

\date{\today}

\begin{abstract}
The critical dynamics of Ising spin glasses with Bimodal, Gaussian, and Laplacian interaction distributions are studied numerically in dimensions 3 and 4. The data demonstrate that in both dimensions the critical dynamic exponent $z_{\rm c}$, the non-equilibrium autocorrelation decay exponent $\lambda_c/z_{\rm c}$, and the critical fluctuation-dissipation ratio $X_{\infty}$ all vary strongly and systematically with the form of the interaction distribution. 
\end{abstract}

\maketitle

\section{Introduction}
At a continuous transition only a few static exponents are enough to completely describe equilibrium critical behaviour. When dynamic measurements are considered the critical behaviour becomes much richer, with further independent critical quantities having non-trivial critical exponents. 

It has been generally assumed that in a given dimension all Ising Spin Glasses (ISGs) lie in a single universality class.
The Hamiltonian of the ISGs is given by
\begin{equation}
{\cal H} = - \sum_{\langle i,j \rangle} J_{ij} S_i S_j
\end{equation}
where $S_i=\pm 1$ are the usual Ising spins and the nearest-neighbour
couplings $J_{ij}$ are random variables.
We have studied numerically critical dynamic behaviour for ISGs having near neighbour interactions in dimensions 
$d = 3$ and  $4$ with Bimodal, Gaussian, or Laplacian distributions of the couplings. \cite{bernardi:97} We find that the dynamic 
exponents vary strongly and systematically from one distribution to another. 
One possible explanation for this could be that for ISGs the 
universality class depends on the form of the interaction distribution. 

In Section II we introduce the dynamical quantities, whereas in Section III we treat the
important point of the reliable determination of critical temperatures. We then present our results for the
spin glasses in four (Section IV) and three (Section V) space dimensions. Finally, Section VI 
contains our conclusions.

\section{Definitions}
In the field of dynamic critical measurements there is not complete consensus as to a universal convention for indicating protocols or exponents, particularly for the case of spin glasses. We will first define the convention that we will use, following \cite{godreche:02,henkel:04}, and relate it to other standard conventions. We will then summarize dynamic scaling properties. This section leans heavily on the review \cite{calabrese:05}.

Throughout, we will implicitly consider only model A dynamics \cite{hohenberg:77} with single spin Glauber (or heat bath) updates in Ising spin glasses; the total time after quench is refered to as $t$. The basic dynamic protocol (there are potentially many others) consists in quenching the sample at time $t=0$ from $T=\infty$ to a temperature $T$, waiting (carrying out updates) for a time that will be called $s$; then at $t=s$ and still at fixed $T$, either the physical conditions are changed in some way (typically by switching on or switching off a small magnetic field $h$) or the instantaneous spin configuration at time $s$ is simply registered to provide a reference state, without physical conditions being changed. There is then an observation period of further updates during which physical parameters are measured as functions of $t$. For convenience the time difference $t-s$ is also denoted $\tau$. In alternative conventions the waiting time $s$ is labeled $t_{\rm w}$, and $\tau$ is labeled $t$. 

Among limiting conditions that can be profitably studied for large samples are the condition $s=0$ (i.e. measurements start immediately on quenching), or alternatively $s \gg \tau$, a long waiting time condition after which the sample is in "quasi-equilibrium". True equilibrium can be achieved if $s$ is "long enough", a criterion that depends in a non-trivial way on the system, on the temperature $T$ and on the sample size $L$. 

Many dynamic observables can be measured. Here we will concentrate on observations at criticality, $T=T_{\rm c}$, for the moment ignoring the question of how $T_{\rm c}$ is to be estimated. 

A first fundamental definition is that of the dynamic critical exponent $z_{\rm c}$ .
At $T_{\rm c}$, the equilibrium autocorrelation relaxation time (with standard single spin updates)
increases with sample size $L$ as
\begin{equation}
\tau_{auto}(L) \sim L^{z_{\rm c}} 
\label{eq:zdef}
\end{equation}
where $z_{\rm c}$ is the dynamical critical exponent. 
%

The two-time autocorrelation function is defined as 
\begin{equation}
C(t,s) = \frac{1}{N} [  \langle \sum_{j=1}^N \langle S_{j}(s)S_{j}(t) \rangle ]
\label{eq:Cts}
\end{equation}
where $\langle \cdots \rangle$ indicates the average over the thermal noise and
$[ \cdots ]$ the average over the disorder
(in alternative conventions $C$ may also be written as $q$).
The critical scaling relation for $C$ is 
\begin{equation}
C(t,s) = s^{-b}f_{c}(t/s).
\label{eq:Ccrit}
\end{equation}
In the quasi-equilibrium limit where $s \gg \tau$ the critical scaling function $f_{c}(t/s)$ should follow the asymptotic behaviour 
\begin{equation}
f_{c}(t/s) \sim [(t/s)-1)]^{-b}.
\label{eq:Cassy1}
\end{equation} 
Thus in this limit we have
\begin{equation}
C(t,s) \sim \tau^{-b}
\label{eq:qeq}
\end{equation}
(in the alternative convention this is written as $q(t) \sim t^{-x}$ with $x \equiv b$). 
For spin glasses the dynamic scaling relation governing $b$ is \cite{ogielski:85}
\begin{equation}
b = (d-2+\eta)/2z_{\rm c}
\label{eq:bscal}
\end{equation}
where $\eta$ is the static critical exponent.
In the opposite limit when $(t/s) \longrightarrow \infty$ 
\begin{equation}
f_{c}(t/s) \sim (t/s)^{-\lambda_c/z_{\rm c}}
\label{eq:Cassy2}
\end{equation}
or for $s=0$
\begin{equation}
f_{c}(t) \sim t^{-\lambda_c/z_{\rm c}}.
\label{eq:Cassy3}
\end{equation}
$\lambda_c/z_{\rm c}$ is related to the "initial slip" exponent $\theta_c$ (an independent critical exponent  \cite{janssen:89,godreche:02}) through
\begin{equation}
\theta_{c} = d/z_{\rm c} - \lambda_c/z_{\rm c}.
\label{eq:theta}
\end{equation}

The two-time linear autoresponse function is
\begin{equation}
R(t,s)=  [ \delta \langle S_{i}(t)\rangle /\delta h(s) ] \mid_{h=0} ~~ (t>s)
\label{eq:Rdef}
\end{equation}
where $h(s)$ is a time dependent conjugate magnetic field.
The scaling equation for $R$ is
\begin{equation}
T R(t,s) = s^{-(1+a)}f_R(t/s)
\label{eq:Rscal}
\end{equation}
with $a = b$ for critical systems.
The critical scaling function $f_{R}(t/s)$ should follow the asymptotic behaviour
\begin{equation}
f_{R}(t/s) \sim (t/s)^{-\lambda_{R}/z_{\rm c}}
\label{eq:Cassy4}
\end{equation}
when $t/s \longrightarrow \infty$. For short range inital correlations $\lambda_{R} = \lambda_{c} $.

In simulations where the field $h$ is applied at $t=0$ and switched off at $t=s$, the following integrated response is measured at times $t>s$:
\begin{equation}
\rho(t,s)  =  T \int\limits_0^s du R(t,u).
\end{equation}
This integrated response is directly related to the commonly studied thermoremanent magnetization:
\begin{equation}
M_{TRM}(t,s) = h \rho(t,s).
\end{equation}

Of further interest is the fluctuation-dissipation ratio $X$ defined by
\begin{equation}
X(t,s) =TR(t,s)/(\delta C(t,s)/\delta s) = \hat{X}(t/s).
\label{eq:Xdef}
\end{equation}
In the quasi-equilibrium condition $s \gg \tau$ the fluctuation-dissipation theorem holds and $X = 1$.\cite{calabrese:05}
When $t/s \longrightarrow \infty$, $X$ takes a limiting value $X_{\infty}$. For $t \gg s$
the ratio $\rho(t,s)/C(t,s)$ also converges to this limit value.
If the amplitudes $A_{c}$ and $A_{R}$  are defined in the same limit $t \gg s$ by
\begin{equation}
f_{c}(t/s) = A_{c}(t/s)^{-\lambda_{c}/z_{\rm c}}
\end{equation}
and
\begin{equation}
f_{R}(t/s) = A_{R}(t/s)^{-\lambda_{c}/z_{\rm c}}
\end{equation}
then \cite{chatelain:04}
\begin{equation}
X_{\infty} =(A_{R}/A_{c})(\lambda_c/z_{\rm c} - b)^{-1}.
\label{eq:Xinf}
\end{equation}
Finally, the dynamic spin glass susceptibility is measured through
\begin{equation}
\chi_{ne}(t) = {1 \over N} [\langle \sum_{j=1}^N S_j^{\alpha}(t) S_j^{\beta}(t) \rangle^2] 
\label{eq:chi}
\end{equation}
where $\alpha$ and $\beta$ are two replicas of the same system relaxing independently.
The infinite time limit to the dynamic SG susceptibility is the equilibrium SG susceptibility for each size $L$, which 
at criticality increases with $L$ as 
\begin{equation}
\chi_{eq}(L) \sim L^{(2-\eta)}
\label{eq:chieq}
\end{equation}
where $\eta$ is the static critical exponent.
The critical time dependence of the non-equilibrium spin-glass 
susceptibility for large samples after a quench to $T_{\rm c}$ and with no anneal ($s=0$) is \cite{huse:89}
\begin{equation} 
\chi_{\rm ne}(t) \sim t^{(2-\eta)/z_{\rm c}} = t^{h^{*}}
\label{eq:chine}
\end{equation}
where $t$ is the time after quench and $z_{\rm c}$ is again the dynamical critical exponent. For convenience we have introduced an exponent $h^{*} =(2-\eta)/z_{\rm c}$.

Even for a canonical continuous transition such as that of the $2d$ Ising ferromagnet, where the static critical exponents are all known analytically and are rational numbers, the dynamic critical exponents can only be established numerically and have non-trivial values.\cite{godreche:02,calabrese:05} For Ising ferromagnets however, field theory (FT) epsilon expansion estimates give reasonable agreement with numerical dynamic exponent estimates in dimensions $3$ and $2$.\cite{calabrese:05} 

It is now well established that for standard systems that are in the same 
universality class not only the static exponents $\nu, \eta$ etc. but also the dynamic exponents $z_{\rm c}$, $\theta_{c}$ and $X_{\infty}$ are all universal. The numerical data discussed in the following
show that in each dimension 
for ISGs expected to lie in the same universality class
the dynamic exponents 
vary strongly with the form of the interaction distribution.

\section{Ordering temperatures}

In order to obtain accurate and reliable simulation values for the dynamic exponents an {\it a priori} requisite is to have reliable estimates for the ordering temperatures $T_c$. 

High temperature series calculations give $T_c$ estimates which are not subject to finite size corrections and are thus intrinsically reliable, but whose accuracy is limited by the number of known terms in the series.  Series estimates can be extremely precise at high dimension but unfortunately they become progressively more inaccurate as the system dimension drops.\cite{singh:87,klein:91,daboul:04} 

There are a number of different ways in which to obtain estimates of ordering temperatures in ISGs through simulations; for most of them one or more critical exponent estimates are also obtained simultaneously. Equilibrium simulations can provide accurate data but necessarily on samples which are of small or moderate size $L$, and the reliability and precision of the $T_c$ estimates are finally limited by the need to extrapolate to large $L$ to eliminate corrections to finite size scaling whose importance depends on the system being studied, on the parameter being measured, and on the maximum range of sample sizes that can be equilibrated with the computing facilities available.  Corrections to scaling are subtle even for the canonical Ising ferromagnets \cite{hasenbusch:99,salas:00,caselle:02} where the leading corrections include both "irrelevant operator" and "analytic" contributions, 
while for ISGs basic guide-lines are lacking so that one must rely on empirical observations. 
In principle it should be possible to use the onset of deviations from strict critical behaviour to monitor $T_c$; for instance $\ln(\chi(L,T))$ against $\ln(L)$ curves bend downwards/upwards for $T$ greater/less than $T_c$. In the SG context this approach has been rarely used as the upbending below $T_c$ is weak, at least in 3d. The equilibrium finite size scaling simulation techniques, which are most often relied on to estimate $T_c$ in ISGs, are based on measurements of parameters that are dimensionless and take on an $L$-independent value at $T_c$ for large $L$. Well known examples are the Binder moment ratio \cite{kawashima:96}  $g(L,T)$ and the second moment correlation length ratio \cite{palassini:99} $\xi(L,T)/L$. Plots of $g(L,T)$ or $\xi(L,T)/L$ as functions of $T$ for fixed $L$ have a unique "crossing point" at a temperature which is equal to $T_c$ in the limit of large $L$. These methods require strict thermal equilibration at each size $L$; also the exact position of the large $L$ limit crossing point may be masked up to quite large $L$ by corrections to scaling (see for instance \cite{beach:05} for the case of the canonical Ising ferromagnet in three dimensions). 

Simulations become heavier with increasing $d$ simply because for given $L$ the number of spins is $L^d$, but this effect is compensated by the fact that $z_{\rm c}$ tends to drop with increasing $d$.
Furthermore crossing points become better defined at higher $d$, and it turns out that corrections to finite size scaling become weaker as $d$ increases. On balance it is in fact easier to estimate $T_c$ reliably by equilibrium simulations at $d=4$ (and above) than at $d=3$.
 
An alternative simulation technique which we will rely on below is to combine static and dynamic measurements to estimate $T_c$ by consistency.\cite{bernardi:96}  This has the advantage of using two dynamic measurements which do not require equilibration and which have negligible corrections to scaling, together with equilibrium spin glass susceptibility measurements which do require equilibration but which are less sensitive to corrections to scaling than are Binder moment ratio or correlation length ratio measurements. 
A range of putative $T_c$ values $T^{*}$ are chosen, and three measurements are made at each $T^{*}$ :

- the effective dynamic exponent $b(T^{*})$ from large $L$ quasi-equilibrium measurements with $s \gg \tau$ using Eq.~(\ref{eq:qeq}),

- the effective dynamic exponent $h^{*}(T^{*})$ from large $L$ measurements of the dynamic SG susceptibility Eq.~(\ref{eq:chine}), and

- the effective static exponent $\eta(T^{*})$ from equilibrium SG susceptibility finite size scaling measurements, Eq.~(\ref{eq:chieq}).

As these three parameters are linked at $T_c$ through the two exponents $\eta$ and $z_{\rm c}$, Eq.~(\ref{eq:bscal}), Eq.~(\ref{eq:chieq})  and Eq.~(\ref{eq:chine}), there is a consistency condition 
which holds at and only at $T^{*}=T_c$. There are different ways to implement this condition. We can first use the equilibrium and dynamic SG susceptibility results together to obtain a set of values of $z(T^{*})$ at each $T^{*}$. \cite{katzgraber:05} These values are to good precision independent of corrections to scaling, and the set of $z(T^{*})$ extends from above to below $T_c$. Secondly, from the effective $\eta(T^{*})$ and $b(T^{*})$ one can derive a second effective $z$, $z^{+}(T^{*})= (d-2+\eta(T^{*}))/2b(T^{*})$. At $T=T_c$ the consistency condition is simply $z(T^{*})=z^{+}(T^{*})$. The measured values of the parameters $\eta(T^{*})$ and $z(T^{*})$  at this unique temperature correspond to the true critical exponents $\eta$ and $z_{\rm c}$. This method is rather insensitive to corrections to finite size scaling.   

The three ISG distributions which will be considered here are the random Bimodal, Gaussian, and Laplacian near neighbour interaction distributions on (hyper)cubic lattices. The explicit normalized distributions are  
\begin{equation} 
P_{B}(J_{ij}) = [\delta (J_{ij}-J)+\delta (J_{ij}+J)]/2,
\label{binomial}
\end{equation}
\begin{equation}
P_{G}(J_{ij}) = \exp(-J_{ij}^2/2 J^2)/(J\sqrt {2 \pi})
\label{gaussian}
\end{equation}
and
\begin{equation}
P_{L}(J_{ij}) = \exp(-\sqrt{2}\mid J_{ij}/J \mid)/(J \sqrt{2}) 
\label{laplacian}
\end{equation}
respectively.
The distributions are symmetric about zero and are normalized in such a way that $<J_{ij}^2>/J^2=1$.

\section{Dimension 4}

We will first consider explicitly dimension 4. $T_c$ values for a range of interaction distributions including the three cases that concern us here were obtained by simulations using the consistency method.\cite{bernardi:97} For the Bimodal and Gaussian distributions the values were fully consistent with and were as accurate or more accurate than other simulation estimates using alternative simulation techniques.\cite{hukushima:99,young,ney-nifle:98,parisi:96,marinari:99}  No other result appears to have been reported for the Laplacian distribution. Essentially negligible corrections to scaling can be seen in the data for any of the simulation techniques at this dimension; for instance, the Binder parameter crossing points are well defined and appear to be independent of $L$ to within high numerical precision. High temperature series estimates for the Bimodal case \cite{singh:87,klein:91,daboul:04} and for other interaction distributions \cite{daboul:04} are in excellent agreement with the simulation estimates; an overview of the data is given in \cite{campbell:05}. The overall agreement between the complementary approaches means that in $d=4$ the consistency simulation technique \cite{bernardi:96} is validated. Hence the $T_c$ values, together with the associated $\eta$ and $z_{\rm c}$ critical exponent values from the consistency method, can be taken as reliable. 

Non-equilibrium measurements of the two-time autocorrelation function and the two-time linear autoresponse function were made at the temperatures corresponding to the $T_c$ values estimated from the consistency method. 
Large systems containing $20^4$ spins were simulated using the
standard heat-bath algorithm. The systems were prepared initially
in a completely disordered state and then quenched down to $T_c$
at time $t=0$. For the computation of the thermoremanent 
magnetization an external field with strength $h = 0.05$ was
applied between $t=0$ and $t=s$ with $s$ varying from 25 to 400.

\begin{figure}
\centerline{\epsfxsize=3.1in\epsfbox
{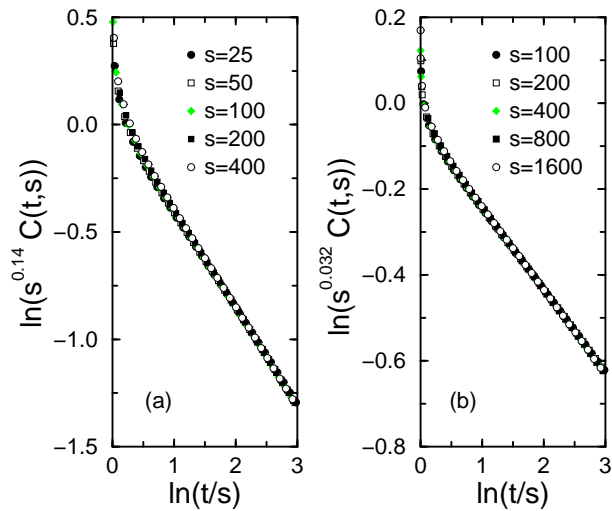}
}
\caption{Scaling behaviour of the autocorrelation function $C(t,s)$
with a Laplacian distribution of the couplings in (a) four and (b)
three dimensions. The best data collapses yield the values of $b$
given in Tables \ref{Table:1} and \ref{Table:2}.
\label{Fig1}}
\end{figure}

\begin{figure}
\centerline{\epsfxsize=3.1in\epsfbox
{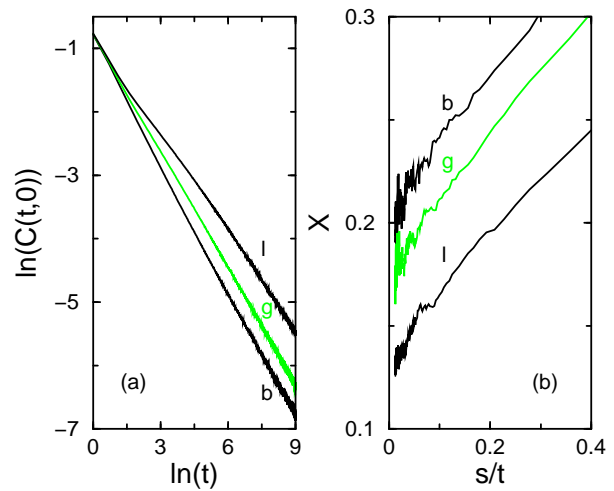}
}
\caption{Out-of-equilibrium quantities in dimension four. 
(a) Time depencence of the autocorrelator with $s=0$. The three
different distributions (b: Bimodal, g: Gaussian, l: Laplacian) yield different
exponents for the power-law behaviour observed at long times. (b) Fluctuation-dissipation
ratio $X$ as function of $s/t$. In the limit $s/t \longrightarrow 0$ one obtains the
limit value $X_\infty$  with different values for the different distributions.
\label{Fig2}}
\end{figure}

Figure \ref{Fig1}a and Figure \ref{Fig2} summarize our findings for the four-dimensional
systems. The expected dynamical scaling behaviour (\ref{eq:Ccrit})
of the autocorrelation function is illustrated in Figure \ref{Fig1}a
for the case of a Laplacian distribution of the couplings.
Plotting $C(t,s)$ as a function of $t/s$ for various values of the
waiting time $s$, an excellent data collapse is achieved for the
value $b=0.140(3)$. Deviations from this scaling behaviour are only
obvious in the regime $t -s \leq s$, i.e.\ outside of the dynamical scaling regime.
A similar good data collapse \cite{footnote} is obtained for the other 
distributions, see Table \ref{Table:1} and Ref. \cite{henkel:04}. It is worth noting
that the values of $b$ obtained in these non-equilibrium simulations
agree with those derived from the quantities 
$\eta$ and $z_c$ via Eq. (\ref{eq:bscal}).
Equilibrium and non-equilibrium simulations therefore consistently 
yield for ISGs critical quantities depending on the form of the
distribution of the couplings.

In Figure \ref{Fig2} we discuss truly non-equilibrium quantities which can not 
be expressed solely by equilibrium quantities. As shown in Figure \ref{Fig2}a
plotting $\ln C(t,0)$ versus $\ln t$ results in straight lines in the
long time limit, in agreement with the expected power-law behaviour (\ref{eq:Cassy3}).
The slopes of these lines yield the exponent $\lambda_c/z_c$. 
Again, this quantity, supposed to be universal, shows a clear dependence
on the chosen distribution, see Table \ref{Table:1}. Finally, 
Figure \ref{Fig2}b displays the temporal evolution of the 
fluctuation-dissipation ratio (\ref{eq:Xdef}) which in the limit $t/s \longrightarrow
\infty$ yields the limit value $X_\infty$, again supposed to be universal.
It is obvious from this plot that the value of $X_\infty$ is different for
the three distributions considered in this work. 

Values for the various parameters corresponding to the present three distributions are shown in Table \ref{Table:1}. The amplitude ratio $A_R/A_c$ 
has been derived from Eq. (\ref{eq:Xinf}).

\begin{table}
\caption{\label{Table:1} Parameter estimates in dimension 4.} 
\begin{tabular}{cccc}
parameter & Bimodal & Gaussian & Laplacian\\ 
\hline
$T_c$ & 2.00(1) [1] & 1.77(1) [1]& 1.53(2) [1]\\ 
$z_{\rm c}$ & 4.45(10) [1] & 5.1(1) [1] & 6.05(10) [1] \\ 
$\eta$ & -0.31(1) [1]& -0.47(2) [1]& -0.60(3) [1] \\ 
$b$ &  0.180(5) [2] & 0.171(2) [28] & 0.140(3) \\ 
$\lambda_c/z_c$ & 0.615(1) [2] & 0.58(1) [28]  & 0.54(1) \\ 
$X_\infty$ & 0.20(1) [2]  & 0.175(10) [28] & 0.13(1) \\
$\theta_c$ & 0.28(2) & 0.205(20) & 0.12(2) \\
$A_{R}/A_{c}$ & 0.087(7) & 0.072(6) & 0.052(6) \\
\hline
\end{tabular}
\end{table}

By inspection of the results in Table I it can be seen that the equilibrium critical exponent $\eta$ together with all the dynamic critical exponents vary strongly from one distribution to another. Apparent non-universality of critical exponents obtained from simulation data has in the past been ascribed to a consequence of errors in the estimation of critical temperatures or to a lack of care in allowing for corrections to finite size scaling. In the present case, the values of the ordering temperatures \cite{bernardi:97} have been validated by high temperature series calculations \cite{daboul:04} and internal evidence, inherent to the consistency method described in Section III, shows that corrections to finite size scaling are negligible in this dimension. In addition, the dynamic parameters are obtained from simulations on large samples and can be considered virtually free of finite-size corrections to scaling. The values of each of the dynamical critical parameters are insensitive to the precise value of the ordering temperature so even if there were small errors in the assumed values of the ordering temperatures the effects on the dynamic critical parameter estimates would be negligible. 

At this point some remarks on the reliability of the extraction of critical exponents from out-of-equilibrium
simulations are in order. The whole approach is based on the assumption that, in the time range in which the
exponents are determined, the dynamical correlation length $\xi(t)$ increases as a simple power-law
\begin{equation} \label{xi1}
\xi(t) \sim t^{1/z_c}
\end{equation}
where $z_c$ is the dynamical critical exponent. For critical ferromagnets this growth law and the resulting
dynamical scaling set in already after a few time steps. However, as spin glasses are characterized by a large value
of $z_c$ one may wonder whether this simple growth low prevails for the times we have accessed in our
simulations or whether a more general growth law of the form
\begin{equation} \label{xi2}
\xi(t) = a t^{1/z_c} + b t^{1/z'}
\end{equation}
with a sizeable finite-time correction is observed.

In fact, the growth of the dynamical correlation length at the critical point of different
three- and four-dimensional spin glasses has been intensively investigated in the recent past.
We mention here the Ising spin glass with a  Bimodal\cite{yoshino:02,berthier:02} or a 
Gaussian\cite{berthier:02,katzgraber:05} distribution of the couplings, the gauge glass with
a Gaussian distribution\cite{katzgraber:05}, the $XY$ spin glass with a Bimodal distribution\cite{yamamoto:04}
or the Heisenberg spin glass with a Gaussian distribution\cite{berthier:04}. All these studies reveal that for times
$t \geq 20$ the increase of the dynamical correlation length in various spin glasses (including
some of the cases we consider in this work) is given by the
simple power-law (\ref{xi1}). No finite-time corrections of the form (\ref{xi2}) have been observed.
Additional support for the growth law (\ref{xi1}) comes from the
perfect dynamical scaling behaviour of two-time quantities (as for example the autocorrelation
function shown in Figure \ref{Fig1}), as a sizeable finite-time
correction would completely spoil the observed data collapse. 

 
\section{Dimension 3}
In dimension $d=3$ the overall situation is rather less satisfactory; high temperature series values \cite{singh:87,klein:91} become imprecise for the Bimodal case and none have been reported for the other distributions. The Binder ratio method becomes delicate because the $g_{L}(T)$ curves only fan out weakly at low temperatures making the limiting crossing point difficult to identify and very sensitive to corrections to scaling.\cite{kawashima:96} The correlation length ratio appears to suffer from strong corrections to scaling, especially at low $L$.\cite{katzgraber:05b} 
For the Gaussian distribution there is a general consensus as to the value of $T_c$ from different simulation estimates.\cite{marinari:98,mari:01,katzgraber:05b} For the Bimodal distribution, published $T_c$ estimates are much more scattered \cite{ogielski:85,kawashima:96,palassini:99,ballesteros:00,mari:02} which we ascribe to difficulties related to corrections to finite size scaling. For the Laplacian distribution we are not aware of other published estimates.

\begin{figure}
\centerline{\epsfxsize=3.1in\epsfbox
{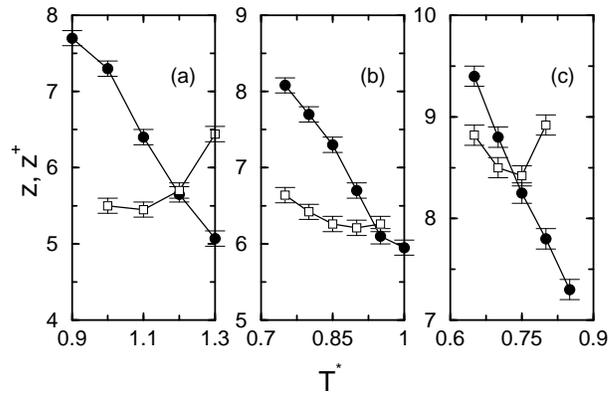}
}
\caption{ Dynamical exponents $z$ (filled circles) and $z^+$ (open squares)
as function of trial temperatures $T^*$ for ISGs in dimension three with
(a) Bimodal, (b) Gaussian and (c) Laplacian distribution of the couplings.
\label{Fig3}}
\end{figure}

We will rely on the value from the consistency method  because, for the reasons outlined above, this technique is much less sensitive to problems of corrections and because it has given excellent agreement with the high temperature series values in $d=4$. 
Data from the consistency method are presented in Figure 3.  With sets of trial temperatures $T^{*}$  we plot   for each system $z(T^{*})$ and $z^{+}(T^{*})$ against $b(T^{*})$.  $z(T^{*})$ is derived from a comparison of the equilibrium and dynamic SG susceptibility results \cite{katzgraber:05} at each $T^{*}$, and $z^{+}(T^{*}) = (d-2+\eta(T^{*}))/2b(T^{*})$. $b(T^{*})$ is measured from the autocorrelation function decay in quasi-equilibrium as defined above, and the effective $\eta(T^{*})$ is obtained from equilibrium finite size SG susceptibility measurements. At $T_c$ consistency of the various exponents dictates that $z(T^{*}) \equiv z^{+} (T^{*})$. The values of $T_c$ together with the exponents $z_{\rm c}$ and $\eta$ obtained are given in Table II (the values are more precise than those given using the same method in \cite{bernardi:96} because of improved equilibrium susceptibility data\cite{palassini:05,katzgraber:05b}).
It is interesting to note that Migdal-Kadanoff estimates of $T_c$ for the different distributions with $d=3$ and the MK parameter $b=2$ \cite{prakash:97,nogueira:99} are strikingly similar to the values given here.

\begin{figure}
\centerline{\epsfxsize=3.1in\epsfbox
{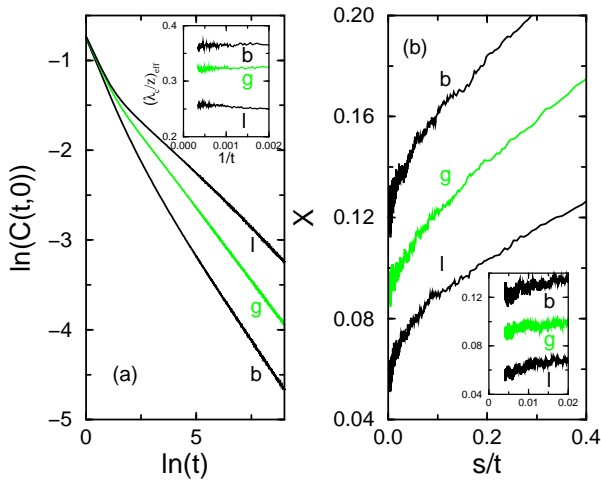}
}
\caption{The same as Figure 2, but now in dimension three.
Whereas the insert in (a) shows the time dependence of the effective exponent 
$(\lambda_c/z_c)_{eff}$ (\ref{eq:effexp}), the insert in (b) displays the long time
behaviour of $X$. 
\label{Fig4}}
\end{figure}

For the non-equilibrium simulations in three dimensions we considered systems with $50^3$ spins
and waiting times $s \leq 1600$. The expected scaling behaviour of the two-time
quantities is again observed, see Figure \ref{Fig1}b for the autocorrelation of the Laplacian
distribution. As shown in Figure \ref{Fig4} (see Table \ref{Table:2})
one observes that the values of $\lambda_c/z_c$ and $X_\infty$
also depend in three dimensions on the form of the distribution function of the couplings.
The insert in Figure \ref{Fig4}a displays the effective exponent
\begin{equation} \label{eq:effexp}
(\lambda_c/z_c)_{eff}  = - (\ln(C(20 t,0)- \ln(C(t, 0))/(\ln(20 t) - \ln(t)) 
\end{equation}
as a function of $1/t$. In all cases this effective exponent 
rapidly reaches a constant, distribution dependent, value.

\begin{table}
\caption{\label{Table:2} Parameter estimates in dimension 3.}
\begin{tabular}{cccc}
parameter & Bimodal & Gaussian & Laplacian\\ 
\hline
$T_c$ & 1.19(1) & 0.92(1)& 0.72(2)\\ 
$z_{\rm c}$ & 5.7(2) & 6.2(1)& 8.6(2)\\ 
$\eta$ & -0.22(2) & -0.42(2)& -0.55(2) \\ 
$b$ &  0.056(3) [2]& 0.043(1) [2]& 0.032(2) \\ 
$\lambda_c/z_{\rm c}$ & 0.362(5) [2]& 0.320(5) [2]& 0.259(2) \\ 
$X_\infty$ & 0.12(1) [2]& 0.09(1) [2]& 0.055(2) \\
$\theta_{\rm c}$ & 0.165(25) & 0.165(13) & 0.090(10) \\
$A_{R}/A_{c}$ & 0.037(4) & 0.025(4) & 0.0125(7) \\
\hline
\end{tabular}
\end{table}

Once again the estimates of the various static and dynamic critical exponents vary considerably from distribution to distribution. The sense of the variations is systematic and is the same as in dimension $4$ :  with increasing kurtosis of the distribution, $T_c$ drops, $z_{\rm c}$ increases, $\eta$ becomes more negative, and the dynamic exponents either expressed as $\lambda_c/z_{\rm c}$ and $X_\infty$ or as $\theta_c$ and $A_{R}/A_{c}$ all drop.

\section{Discussion}

For standard continuous transitions the renormalization group theory provides a comprehensive explanation of critical behaviour and in particular of the strict identity of exponents of all systems within each universality class, the class being defined by a restricted list of parameters which includes the physical dimension $d$ and the number of order parameter components $n$. The universality covers not only equilibrium exponents but extends to the whole family of dynamic exponents. This universality reflects the fundamental principle that within each class, the details of the physics at the local level do not affect the large scale behaviour which determines the critical exponents. 

It has been widely assumed that in a given dimension all ISGs fall in the
same universality class. However it should be noted that the critical
behaviour of spin glasses is qualitatively very different from that of
standard systems such as ferromagnets. The upper critical dimension is 6
rather than 4, and below the upper critical dimension the 
specific heat exponents are strongly negative so there is no specific heat
peak or cusp. Field theory (which provides the well known $\epsilon$
expansion development at standard continuous transitions) has proved
intractable in the ISG context below the upper critical dimension $d=6$.
\cite{dedominicis:98} Already at $d=5$ and $d=4$, numerical values of the
equilibrium exponents obtained from summing the known leading terms to
order three in the ISG $\epsilon$ expansion \cite{yeo:05} are very different
from estimates using the high temperature series method
\cite{klein:91,daboul:04} or simulations. For the Ising ferromagnet at
$d=3$ and so $\epsilon=1$, the FT development in $\epsilon$ to third order
is accurate to better than $0.001$.\cite{zinn-justin:89} This is in total
contrast to the situation for the Bimodal ISG where at $d=5$ (so again at
$\epsilon = 1$) the FT sum to third order in $\epsilon$ gives $\eta(d=5) =
1.6897$, strikingly different from the high temperature series value \cite{klein:91}
$\eta(d=5) = -0.38(7)$ and the simulation estimate
$\eta(d=5) = -0.39(2)$.\cite{bernardi:97} This implies that a sum
including many further terms (oscillating in sign) would be needed to
finally obtain stable and accurate FT predictions. In practice,
establishing such a sum seems entirely ruled out, but the question remains
open as to whether the necessary quasi-cancellations among the unknown
higher order FT terms could depend on parameters such as the lattice
structure or the form of the interactions.

In our simulations of the Ising spin glasses we have found that,
in most cases, the differences between the exponent estimates for the different systems are much larger than the statistical error bars. 
This is especially obvious for the amplitude ratio $A_R/A_c$ which is supposed to yield clearly distinct values for
different universality classes, similar to what is observed for static ampltude ratios at equilibrium.
Extrapolations in each dimension suggest that if the interaction distribution is modified and $T_c$ decreases,  all the exponents studied vary systematically in such a way that $z_{\rm c}$ increases strongly, $\eta$ tends towards a value near $2-d$ (which is the strict limiting value for $\eta$ in each dimension when $T_c = 0$) while  $X_\infty$ and $\theta_c$ tend to near zero. The data as they stand are thus compatible with exponents each varying continuously towards a $T_c=0$ limit as $T_c$ is driven lower by a widening of the interaction distribution (increasing kurtosis).

Our non-equilibrium simulations can in principle be subjected to only three different sources of systematic errors:
(1) the values of the critical temperatures are erroneous, (2) the sizes of the samples are too small, or (3)
the time range of our runs is insufficient. Let us address these three different points.

The values of the critical temperatures we use have been estimated with a technique combining equilibrium
and non-equilibrium measurements\cite{bernardi:97} and are in excellent agreement with independent simulation
estimates. In addition, $T_c$ in the four-dimensional systems have been confirmed recently\cite{daboul:04}
by high temperature series estimates. This agreement validates the technique used for the evaluation of $T_c$ and
also, as the method relies on a consistency argument, indirectly the values of $z_c$ and $\eta$ obtained in the
same simulations.

The data we have discussed in this paper concern principally dynamic exponents in ISGs. 
These measurements have the advantage of not requiring strict equilibration, which in turn 
permits studies on samples so large that they cannot be conveniently equilibrated at or near criticality.  
As the sample sizes $L$ are much larger than the maximum of the correlation length $\xi(t)$
attained during the simulations, the measurements are always taken in the infinite sample size 
limit and are not hampered by finite-size effects.
We have thus been able to minimize one of the major sources of systematic errors in numerical
simulations, namely corrections to finite size scaling.
This is very similar to what is observed when studying critical ferromagnets, as for
example the three-dimensional Ising model \cite{jaster:99}, where no notable finite-size corrections to scaling are encountered
in non-equilibrium simulations,
in contrast to equilibrium simulations where these corrections to scaling can be very strong.

Whereas finite-size effects are well controlled in non-equilibrium simulations
of large systems, this is not immediately obvious for finite-time corrections.
Indeed, for spin glasses the critical dynamical exponent $z_c$ takes on very large values, 
whch might raise some doubts whether the simple power-law increase (\ref{xi1}) of the dynamical
correlation length is valid in the time range we accessed.
However, various investigations of spin glasses in different dimensions and with different distributions of the couplings
in the recent past have found that in general the increase of $\xi(t)$ is completely described by (\ref{xi1}) for times $t \geq 20$.
We can therefore be confident that the dynamical scaling approach,
underlying our estimates of the critical quantities, is also valid for spin glasses and that 
the values of the equilibrium and non-equilibrium critical quantities we obtain are reliable. 
Furthermore, as shown in Figure \ref{Fig4}, the run times of our simulations are clearly sufficient
for the dynamic parameters to take up their limiting values.

Having now excluded the most probable sources of systematic errors we interpret our numerical data as
strong evidence that in spin glasses critical quantities do depend on the exact form of the  
distribution of the couplings. Of course, as we only provide numerical evidence, 
we can not completely exclude that corrections coming from other sources could have some impact on the values of the
critical exponents.

\acknowledgments
We would like to thank H.G. Katzgraber for useful discussions and for making his data available, and M. Palassini for kindly providing equilibrium susceptibility data.
MP acknowledges the support by
the Deutsche Forschungsgemeinschaft through grant no. PL 323/2. 
The non-equilibrium simulations were done on the IBM supercomputer Jump at the NIC J\"{u}lich
(project Her10). 
Complementary simulations
were done by J.\ Kirmair.


\begin{references}
\bibitem{footnote} In the case of the Gaussian distribution in four dimensions the
values listed in Table \ref{Table:1} correct those given in Reference \cite{henkel:04}
which were obtained with an erroneous code.
\end{references}
\end{document}